\documentclass[aps,prd,showpacs,
]{revtex4}
\usepackage{graphicx}
\usepackage{hyperref}
\usepackage{amssymb, amsmath}

\def\la{\; \raise0.3ex\hbox{$<$\kern-0.75em\raise-1.1ex\hbox{$\sim$}}\;}
\def\ga{\;  \raise0.3ex\hbox{$>$\kern-0.75em\raise-1.1ex\hbox{$\sim$}}\;}

\def\pFn{p_{\raise-0.3ex\hbox{{\scriptsize F$\!$\raise-0.03ex\hbox{\rm n}}}}
}  % p_Fn
\def\pFp{p_{\raise-0.3ex\hbox{{\scriptsize F$\!$\raise-0.03ex\hbox{\rm p}}}}
}  % p_Fp
\def\pFe{p_{\raise-0.3ex\hbox{{\scriptsize F$\!$\raise-0.03ex\hbox{\rm e}}}}
}  % p_Fe
\def\pFmu{p_{\raise-0.3ex\hbox{{\scriptsize F$\!$\raise-0.03ex\hbox{
\rm $\mu$}}}} }  % p_Fe
\def\m@th{\mathsurround=0pt }
\def\eqalign#1{\null\,\vcenter{\openup1\jot \m@th
   \ialign{\strut$\displaystyle{##}$&$\displaystyle{{}##}$\hfil
   \crcr#1\crcr}}\,}
\def\dd{\mbox{d}}

\begin{document}

%%%%%%%%%%%%%%%%%%%%%%%%%%%%%%%%%%%%%%%%%%%%%%%%%%%%%%%%%%%%%%%%%%%%%%
\title{Neutrino Emissivity of Non-equilibrium beta processes With Nucleon Superfluidity}
\author{Chun-Mei Pi, Xiao-Ping Zheng,\footnote{zhxp@phy.ccnu.edu.cn} Shu-Hua Yang}
\affiliation{Institute of Astrophysics, Huazhong Normal University,
Wuhan 430079, China}
\date{\today}
%

%\preprint{}
\pacs{
97.60.Jd,    %Neutron stars
95.30.Cq,    %Elementary particle processes
26.60.Dd,    %Neutron star core
23.40.Bw     % Weak-interaction and lepton aspects
}

%%%%%%%%%%%%%%%%%%%%%%%%%%%%%%%%%%%%%%%%%%%%%%%%%%%%%%%%%%%%%%%%%%%%%%
\begin{abstract}
We investigate the influence of nucleon superfluidity on the
neutrino emissivity of nonequilibrium $\beta$ processes.
Calculations of the reduction factors for direct and modified Urca
processes with three types of nucleon superfluidity in $npe$ matter
are performed. The numerical results are given because the
analytical solution is impossible. We find that the superfluid
influence is closely related to the chemical departure from $\beta$
equilibrium. For a small chemical departure, the superfluid
reduction factor depends almost only on the gap and is hardly
affected by the departure, while for a large enough departure, it
rapidly enhances neutrino emissivity. The onset of "enchanced"
emission has some corresponding thresholds that seem to be linked to
the ratio of the energy gap to the chemical departure.

\end{abstract}

\pacs{97.60.Jd, 95.30.Cq, 26.60.Dd, 23.40.Bw}

\maketitle

%%%%%%%%%%%%%%%%%%%%%%%%%%%%%%%%%%%%%%%%%%%%%%%%%%%%%%%%%%%%%%%%%%%%%%%%%%%%%
%**************** Section 1 ******************************
\section{Introduction}
\label{1}
%%%%%%%%%%%%%%%%%%%%%%%%%%%%%%%%%%%%%%%%%%%%%%%%%%%%%%%%%%%%%
In superdense nuclear matter that constructs the core of a neutron
star, a great amount of energy is carried away by the neutrino. For
the standard composition of superdense matter (neutrons, with an
admixture of protons and electrons), the main neutrino emission
mechanisms are powerful direct Urca reactions,
\begin{eqnarray} \label{eq:Durca}
n \rightarrow p + e + \overline{\nu_e} ,\quad
p + e \rightarrow n  + \nu_e ,
\end{eqnarray}
and much weaker modified Urca reactions,
\begin{eqnarray} \label{eq:Murca}
n + N \rightarrow p + N + e + \overline{\nu_e} , \quad
p + N + e \rightarrow n + N + \nu_e .
\end{eqnarray}
The direct Urca reactions can proceed only if the ratio of the
proton number density to the total baryon number density $n_{p}/n$
exceeds a certain threshold value which allows simultaneous energy
and momentum conservation \cite{lpph91}. Otherwise, direct Urca
reactions are forbidden. The most powerful neutrino energy losses
are produced by modified Urca reactions.

Furthermore, the $\beta$ reactions just mentioned bring the
constituents into the state of chemical equilibrium, $\mu_{n} =
\mu_{p} + \mu_{e}$, which determines the relative concentrations of
particles. However, if any external or macroscopic phenomenon
changes the density of a matter element, the $\beta$ equilibrium
state will change, resulting in a departure from equilibrium
quantified by the chemical imbalance $\delta\mu = \mu_{n} - \mu_{p}
- \mu_{e}$. Several authors have investigated many astrophysical
situations in which the nonequilibrium $\beta$ processes occur, such
as gravitational collapse of neutron star \cite{h92}, pulsar
spin-down \cite{rei95, rei97, fr05}, a hypothetical time variation
of the gravitational constant \cite{jrf06}, the existence of even
relatively slow hydrodynamic flow \cite{us96} or of millisecond
oscillations \cite{rg92} in the neutron star interiors. The
departure from equilibrium, $\delta\mu \neq 0$, increases the phase
space for nonequilibrium $\beta$ process and strongly enhances
corresponding neutrino emissivity with respect to the chemical
equilibrium value \cite{h92}.

It is well known that below a certain critical temperature nucleons
in neutron star matter will be in superfluid states. The appearance
of energy gaps $\Delta_{n}$ and $\Delta_{p}$ reduces the particle
momentum space, which contributes to the reaction rates; that is,
the "effective" widths of momentum space of the reacting particles
near the Fermi surfaces become thinner. This suppresses the reaction
rate and the neutrino emissivity. In $\beta$-equilibrium matter,
superfluid reductions in reaction rate and neutrino emissivity are
dependent on the ratio of the gap to the temperature $\Delta/T$. For
strong superfluidity, the reductions decay exponentially, while for
intermediate values of $\Delta/T$, the situation is more
sophisticated. The detailed expressions are given in a
review\cite{ykgh01}.

The departure from equilibrium $\delta\mu$ increases the neutrino
emissivity, while the superfluid energy gaps behave in the opposite
way. Indeed, the nonequilibrium $\beta$ reaction with nucleon
superfluidity is a new case that is worth studying. Reisenegger
\cite{rei97} first analyzed the mechanism to produce a faster
reaction owing to nonequilibrium with neutron and proton
superfluidity. Using a crude model, he assumed that for $\delta\mu <
\Delta_p + \Delta_n$, $\beta$ reactions were completely suppressed,
while for $\delta\mu
> \Delta_p + \Delta_n$ the effect of superfluidity could be
neglected. Villain and Haensel broke through Reisenegger's steplike
modeling and calculated the precise reduction factors of the net
reaction rates for nonequilibrium direct and modified Urca processes
in the presence of various types of superfluidity by means of
sophisticated numerical methods \cite{vh05}. However, neutrino
emissivity is also an important quantity. We need to calculate the
relevant neutrino emissivities, which, along with the net reaction
rates, can be used to simulate the evolution of superfluid neutron
star cores that are off $\beta$ equilibrium. Our primary goal here
is to obtain the total neutrino emissivities of nonequilibrium
$\beta$ processes in superfluid circumstances.

The paper is organized as follows. In Sec. II, we recall the
nonequilibrium condition and basic features of nucleon
superfluidity. In Secs. III and IV, we detail the numerical
calculations of neutrino emissivities for Durca and Murca reactions
and show our results. Section V includes a short summary of our
conclusions and discussion of possible applications.

%%%%%%%%%%%%%%%%%%%%%%%%%%%%%%%%%%%%%%%%%%%%%%%%%%%%%%%%%%%%%%%%%%%%%%%
\section{Description of chemical derivation and nucleon superfluidity}
\label{2}
%%%%%%%%%%%%%%%%%%%%%%%%%%%%%%%%%%%%%%%%%%%%%%%%%%%%%%%%%%%%%%%%%%%%%%%
We consider a simple $npe$ model of neutron star matter. Each of the
constituents is strongly degenerate, with Fermi momenta $p_{F_{i}}$
and chemical potential $\mu_{i}$ ($i=n,p,e$). They are in
thermodynamic equilibrium but not necessarily in chemical
equilibrium. In the absence of $\beta$ equilibrium, there is a
finite difference in the chemical potentials, $\delta\mu = \mu_{n} -
\mu_{p} - \mu_{e} \neq 0$, following the conventions of Ref.
\cite{h92}.

Nucleon superfluidity occurs via Cooper pairing of particles owing
to an attractive component of their interaction, with the appearance
of a gap $\delta$ in the particle energy spectrum near the Fermi
level \cite{ls01}. It is widely accepted that there are likely three
types of nucleon superfluidity in $npe$ matter, $^1$$S$$_0$,
\mbox{$^3$$P$$_2$ ($m_J=0$)} and \mbox{$^3$$P$$_2$ ($|m_J|=2$)},
denoted A, B, and C, respectively (Table \ref{tab:ABC}) \cite{ly94}.
We introduce two dimensionless quantities, $\xi=\delta\mu/T$ and
$v=\Delta(T)/ T$, to describe the chemical deviation and gap
amplitude, for convenience, and use natural units with $\hbar =
k_{B} =c=1$ throughout this paper.

To guarantee the integrity of this paper, we recall fundamental
properties and characteristic quantities
\cite{lp80,tamagaki70,yls99}. The onset of superfluidity is
accompanied by the appearance of the energy gap $\delta$. Near the
Fermi surface ($|p-p_{F}|\ll p_{F}$), we have
\begin{equation} \label{eq:Gap}
\begin{array}{l}
      \varepsilon\,=\,\mu\,-\,\sqrt{\delta^2 + \eta^2}
      \; \; {\rm at} \; \; p < p_{\rm F} \, ,
      \quad
      \varepsilon = \mu + \sqrt{\delta^2 + \eta^2}
      \; \; {\rm at} \; \; p \ge p_{\rm F} \;.
\end{array}
\end{equation}
Here $\eta=v_{\rm F}(p-p_{\rm F})$, $p_{\rm F}$ and $v_{\rm F}$ are
the Fermi momentum and Fermi velocity, respectively; $\mu$ is the
chemical potential; and $\delta^2=\Delta^2 (T)F(\vartheta)$, where
$\Delta(T)$ is an amplitude that determines the temperature
dependence of the gap and $F(\vartheta)$ describes the dependence of
the gap on the angle $\vartheta$ between the quantization axis and
the particle momentum. The quantities $\Delta$ and $F$ are
determined by the superfluidity type (Table \ref{tab:ABC}). In case
A the gap is isotropic, and $\delta = \Delta(T)$. In cases B and C,
the gap depends on $\vartheta$. Note that in case C the gap vanishes
at the poles of the Fermi sphere at any temperature: \mbox{$ F_{\rm
C}
(0) = F_{\rm C} (\pi) = 0$}.\\
\begin{table}
\renewcommand{\arraystretch}{1.2}
\caption{Three types of superfluidity }
\begin{center}
  \begin{tabular}{||c|cccc||}
  \hline \hline
Type\rule{0em}{2.5ex}
         & Superfluidity type &$\lambda$&   $F(\vartheta)$
         & $T_c/\Delta(0)$  \\
  \hline
  A       & $^1{\rm S}_0$    &  1&  1                     & 0.5669
                                                        \rule{0em}{3ex} \\
  B       & $^3{\rm P}_2\ (m_J =0)$ &  1/2
                             & $(1+3\cos^2\vartheta)$ & 0.8416            \\
  C       & $^3{\rm P}_2\ (|m_J| =2)$ &  3/2
                             & $\sin^2 \vartheta$     & 0.4926  \\
  \hline \hline
\end{tabular}
\label{tab:ABC}
\end{center}
\end{table}
The gap amplitude $\Delta(T)$ is determined by the BCS theory which can be written as
\begin{equation}\label{eq:BCS}
\ln\frac{\Delta_{0}}{\Delta(T)}=2\lambda\int\frac{d\Omega}{4\pi}\int_{0}^{\infty}\frac{dx}{z}fF(\vartheta),
\end{equation}
where $\Delta_{0}=\Delta(0)$, $d\Omega$ is a solid angle element in
the direction of particle momentum $\mathbf{p}$, $f=(1+e^{z})^{-1}$
is the Fermi-Dirac distribution, $\lambda$ is a numeric coefficient
(Table \ref{tab:ABC}), and
\begin{equation}\label{eq:dimen}
z=\frac{\varepsilon - \mu}{T}=sign(x)\sqrt{x^{2}+y^{2}}, \quad  x=\frac{\eta}{T}, \quad  y=\frac{\delta}{T}
\end{equation}
Note that in the absence of nucleon superfluidity, $z=x$. According
to Table \ref{tab:ABC}, the dimensionless gap $y$ can be written as:
\begin{equation}
     y_{\rm A} = v_{\rm A}, \quad y_{\rm B} =
     v_{\rm B} \, \sqrt{1+3\cos ^2\vartheta},
     \quad
     y_{\rm C} = v_{\rm C} \, \sin \vartheta\,.
\label{y}
\end{equation}

In the following calculations that involve nucleon superfluidity, we
assume that only one type of nucleon is superfluid, which is
equivalent to assuming that the larger gap prevails. This
approximation seems to be quite reasonable \cite{ls01}. Our goal in
this paper is to investigate the influence of nucleon superfluidity
on the neutrino emissivity of nonequilibrium $\beta$ processes,
which is described by the reduction factor $R^i_{X}(\xi,v_j)=
Q_{X}(\xi,v_j) / Q_{X}(\xi)= I^i_{X}(\xi,v_j) / I^i_{X}(\xi)$.
$Q_{X}(\xi,v_j)$ and $Q_{X}(\xi)$ refer to the neutrino emissivities
in two cases: one is nonequilibrium $\beta$ process with nucleon
superfluidity; the other is nonequilibrium $\beta$ process without
nucleon superfluidity. Furthermore, $X$ labels the type of
reactions, $i$ denotes the superfluidity type (A, B, or C), and $j$
the type of superfluid nucleon ($n$ or $p$) with gap $v_j$. Note
that for $T > T_c$ (the critical temperature of the superfluid
nucleon), $R^i_{X}(\xi,v_j) = R^i_{X}(\xi,0) = 1$, and for $T <
T_c$, $R^i_{X}(\xi,v_j) < 1$.
%%%%%%%%%%%%%%%%%%%%%%%%%%%%%%%%%%%%%%%%%%%%%%%%%%%%%%%%%%%%%%%%%%%%%%%
\section{Direct Urca processes}
\label{3}
%%%%%%%%%%%%%%%%%%%%%%%%%%%%%%%%%%%%%%%%%%%%%%%%%%%%%%%%%%%%%%%%%%%%%%%
As shown in Ref. \cite{lpph91}, the direct Urca process is allowed
by the momentum conservation when $p_{F_{n}}<p_{F_{p}}+p_{F_{e}}$.
For pure $npe$ matter where $p_{F_{p}} = p_{F_{e}}$, it corresponds
to $n_{p} / n > 1/9$. This happens if the density is several times
higher than the standard nuclear matter density,
$\rho_{0}=2.8\times10^{14}$ g cm$^{-3}$.

In the absence of $\beta$ equilibrium, the neutrino emissivities of
two direct Urca processes with nucleon superfluidity are in the
following forms:
\begin{eqnarray}
&&Q_n^{(D)}(\xi,v_j)=\frac{4\pi}{(2\pi)^{8}} T^6\left[ \prod_{j=1}^3
\int d\Omega_j\right]\;
\delta(\mathbf{P}_f-\mathbf{P}_i)|M_{fi}|^{2}\prod_{j=1}^3 p_{F_j}
m_j^\ast \int_0^\infty d x_\nu \; x_\nu^3
\nonumber \\
&&  \int_{-\infty}^{+\infty} d x_1 \; f(z_1)
\int_{-\infty}^{+\infty} d x_2 \;(1-f(z_2))
    \int_{-\infty}^{+\infty} d x_3 \;(1-f(x_3))\delta(z_2+x_3-z_1+x_\nu + \xi),
\label{eq:Qdn}
\end{eqnarray}
\begin{eqnarray}
&&    Q_p^{(D)}(\xi,v_j)=\frac{4\pi}{(2\pi)^{8}} T^6\left[
\prod_{j=1}^3 \int d\Omega_j\right]\;
\delta(\mathbf{P}_f-\mathbf{P}_i)|M_{fi}|^{2} \prod_{j=1}^3 p_{F_j}
m_j^\ast \int_0^\infty d x_\nu \; x_\nu^3
\nonumber \\
&&  \int_{-\infty}^{+\infty} d x_2 \; f(z_2)
\int_{-\infty}^{+\infty} d x_3 \;f(x_3)
    \int_{-\infty}^{+\infty} d x_1 \;(1-f(z_1))\delta(z_1-z_2-x_3-x_\nu - \xi).
\label{eq:Qdp}
\end{eqnarray}
where $j=1,\,2,\,3$ corresponds to $n,\,p,\,e$ respectively, $x_\nu=
p_{\nu}/T=\varepsilon_{\nu}/T$ is the dimensionless energy of the
neutrino, $\xi$, $v_j$, and $z_j\,(j=1,\,2)$ have been defined in
Sec. \ref{2}, and $f(x)=(1+e^x)^{-1}$ is the Fermi-Dirac function of
nucleons and electron, $p_{F_j}$ is the corresponding Fermi momentum
and $m_j^\ast$ is the effective particle mass. Furthermore,
$d\Omega_j$ is the solid angle element in the direction of the
particle the momentum $\mathbf{p}_j$, and the $\delta$ functions
describe momentum and energy conservations of the particles in the
initial and final states. Finally, $|M_{fi}|^2$ is the squared
reaction amplitude, and for nonrelativistic nucleons, it is
independent of particle momenta and can be taken out of the
integral. In this paper, we focus on the total neutrino emissivity
$Q_D=Q_n^{(D)}+Q_p^{(D)}$.

Let us start with the total neutrino emissivity $Q_{0}^{(D)}$ of the
direct Urca processes without nucleon superfluidity under $\beta$
equilibrium. Under the condition of $\beta$ equilibrium and without
nucleon superfluidity, we set $\xi =0$ and replace $z_j$ with $x_j$.
The direct and inverse reactions have the same neutrino emissivity.
Then $Q_{0}^{(D)}$ can be written as (for details see Ref.
\cite{ykgh01}):
\begin{eqnarray}
&&    Q_{0}^{(D)}  =  {2 \over (2 \pi)^{8}} T^6 A_{D}I_{0}^{(D)}|M_{fi}|^{2} \prod_{j=1}^3 p_{F_j} m_j^\ast,
\label{eq:DecompDur} \\
&&   A_{D} = 4\pi\left[ \prod_{j=1}^3 \int d\Omega_j\right]\delta(\mathbf{P}_f-\mathbf{P}_i),
\label{eq:Adur} \\
&&   I_{0}^{(D)} = \int_0^\infty d x_\nu \; x_\nu^3
       \left[ \prod_{j=1}^3 \int_{-\infty}^{+\infty}
       d x_j \; f_j \right]
       \delta \left( \sum_{j=1}^3 x_j-x_\nu \right),
\label{eq:Idur}
\end{eqnarray}
Here the integrals $A_{D}$ and $I_{0}^{(D)}$ are standard (\cite{st83}):
\begin{equation}
     A_{D} =\frac{32\pi^2}{p_{F_n}p_{F_p}p_{F_e}}, \quad I_{0}^{(D)} =\frac{457\pi^6}{5040}\,.
\end{equation}

%%%%%%%%%%%%%%%%%%%
\subsection{Nonequilibrium without superfluidity}
In this section, we recall the neutrino emissivity of nonequilibrium
Durca processes without nucleon superfluidity. When $\beta$
processes are off chemical equilibrium, $\xi \neq 0$, the direct and
inverse reactions of Urca processes have different neutrino
emissivities. The expressions for the neutrino emissivities of the
two reactions are similar to Eqs. (\ref{eq:Qdn}) and (\ref{eq:Qdp}),
but the variable $z_j$ is replaced with the corresponding $x_j$. The
total neutrino emissivity can be presented in the forms
\cite{h92,rei95}
\begin{equation}
Q_{D}(\xi)=\frac{1}{(2\pi)^{8}}T^{6}A_{D}I_{D}(\xi)|M_{fi}|^{2} \prod_{j=1}^3 p_{F_j} m_j^\ast,
\end{equation}
with
\begin{eqnarray}
&& I_{D}(\xi)=\int_0^\infty dx_\nu x_\nu^3 \left[J(x_\nu - \xi) + J(x_\nu + \xi)\right], \\
&& J(x_\nu - \xi)= \left[ \prod_{j=1}^3 \int_{-\infty}^{+\infty}
       d x_j \; f_j \right]\delta(x_1 + x_2 + x_3 - x_\nu + \xi),
\label{eq:Jminus} \\
&& J(x_\nu + \xi)= \left[ \prod_{j=1}^3 \int_{-\infty}^{+\infty}
       d x_j \; f_j \right]\delta(x_1 + x_2 + x_3 - x_\nu -\xi).
\label{eq:Jplus}
\end{eqnarray}

According to Ref. \cite{rei95},
\begin{eqnarray}
&& Q_{D}(\xi)=Q_{0}^{(D)}F_D(\xi), \\
&& F_D(\xi) =1+\frac{1071\xi^2}{457\pi^2}+\frac{315\xi^4}{457\pi^4}+\frac{21\xi^6}{457\pi^6}.
\end{eqnarray}
We obtain
\begin{equation}
I_{D}(\xi)=\frac{457\pi^6}{2520}\left (1+\frac{1071\xi^2}{457\pi^2}+\frac{315\xi^4}{457\pi^4}+\frac{21\xi^6}{457\pi^6} \right).
\end{equation}
%
%%%%%%%%%%%%%%%%%%%
\subsection{Nonequilibrium with superfluidity}
%%%%%%%%%%%%%%%%%%%
Consider suppression of nonequilibrium Direct Urca processes by
proton or neutron superfluidity. The superfluidity affects the
dispersion relation of nucleons under the integrals in Eqs.
(\ref{eq:Jminus}) and (\ref{eq:Jplus}), in accordance with Eq.
(\ref{eq:Gap}). The neutrino emissivity can be written as
\begin{equation}
Q_{D}(\xi,v_j)= Q_{D}(\xi) R^i_{D}(\xi,v_j).
\end{equation}
$R^i_{D}(\xi,v_j)$ is the superfluid reduction factor. The total
neutrino emissivity $Q_{D}(\xi,v_j)=Q_n^{(D)}(\xi,v_j)
+Q_p^{(D)}(\xi,v_j) $ is in the following form:
\begin{equation}
Q_{D}(\xi,v_j)= \frac{1}{(2\pi)^{8}} T^6 A_{D} I_{D}^i(\xi,v_j)|M_{fi}|^{2} \prod_{j=1}^3 p_{F_j} m_j^\ast,
\end{equation}
where
\begin{eqnarray}
&& I_{D}^i(\xi,v_j)=\frac{4\pi}{A_{D}}\left[ \prod_{j=1}^3 \int d\Omega_j\right]\; \delta(\mathbf{P}_f-\mathbf{P}_i)\int_0^\infty dx_\nu x_\nu^3  \,
\nonumber \\
&&               \int_{-\infty}^{+\infty} dx_1 dx_2 dx_3  \,
                \left[ f_1(1-f_2)(1-f_3)\delta(z_1 -z_2-x_3-x_\nu+\xi)  \,
                +f_2f_3(1-f_1)\delta(z_1 +x_\nu-z_2-x_3+\xi)\right], \qquad
\label{eq:IDSF}
\end{eqnarray}
and $d\Omega_j$ is the element of the solid angle in the direction of the momentum of superfluid nucleon $j$.

\begin{figure}[t]
\setlength{\unitlength}{1mm}
\leavevmode
\hskip  0mm
\includegraphics[width=0.5\textwidth]{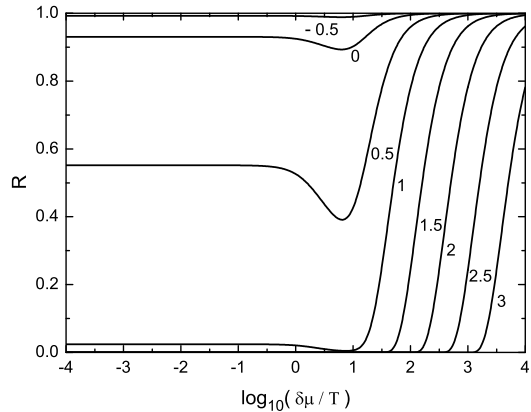}
\caption
{Dependence of the reduction factor on the chemical imbalance for various gap's amplitude. The numbers next to the curves give value of $\log_{10}(gap/T)$. Direct reaction with isotropic superfluidity of protons or neutrons. }
\label{1fig}
\end{figure}

Here, we consider the case where only one type of nucleon is
superfluid, then the $z$ variable for the nonsuperfluid nucleon has
to be replaced with the corresponding $x$ variable, while $v_j$ is
"included" in the $z$ variable for the superfluid nucleon. Thus, the
reduction factor of nucleon superfluidity for a nonequilibrium Durca
process can be presented by
\begin{eqnarray}
&& R^i_{D}(\xi,v_j)= Q_{D}(\xi,v_j) / Q_{D}(\xi)=\frac{I_{D}(\xi,v_j)}{I_{D}(\xi)} \,
\nonumber \\
&&               = \frac{1}{I_{D}(\xi)}\int_0^{\pi/2}\sin(\vartheta)
d\vartheta H_{D}^i(\xi,v_j), \qquad
\end{eqnarray}
where
\begin{eqnarray}
&& H_{D}^i(\xi,v_j)=\int_0^\infty dx_\nu x_\nu^3  \,
\int_{-\infty}^{+\infty} dx_1 dx_2 dx_3  \,
                [ f_1(1-f_2)(1-f_3)\delta(z_1 -z_2-x_3-x_\nu+\xi)  \,
\nonumber \\
&&
                +f_2f_3(1-f_1)\delta(z_1 +x_\nu-z_2-x_3+\xi)]. \qquad
\label{eq:HDSF}
\end{eqnarray}
In Eq. (\ref{eq:HDSF}), integration over the electron variable can
be done using the formula for Fermi integrals, similar to the
treatment in Ref. \cite{vh05},
\begin{equation} \label{Eq:Gy}
\int\limits_{-\infty}\limits^{+\infty} dx f(x)f(y-x)=\frac{y}{e^y-1}\widehat{=}G(y)
\end{equation}
which enables us to integrate over the $x$ variable for the
nonsuperfluid nucleon, giving
\begin{eqnarray} \label{Eq:cut}
&& H_{D}^i(\xi,v_j)= \int\limits_0\limits^\infty dx_\nu  x_\nu^3
       \int\limits_0\limits^{+\infty} dx_j [f(z_j)G(x_\nu-\xi-z_j)+f(-z_j)G(x_\nu-\xi+z_j)\
\nonumber \\
&& +f(z_j)G(x_\nu+\xi-z_j)+f(-z_j)G(x_\nu+\xi+z_j)].
\end{eqnarray}
We assume only one type of nucleon superfluidity and numerically evaluate $R^i_{D}(\xi,v_j)$ for the wide range of interest with algorithmic methods (refer to the appendix of \cite{vh05}). In Fig. \ref{1fig} we present the reduction factor for Durca reaction with isotropic superfluidity of protons or neutrons. The variable is dimensionless chemical departure $\log_{10}(\delta\mu/T)$. The curves correspond to different values of dimensionless gap amplitude $\log_{10}(gap/T)$. From this figure we can get that there are at least two features: (I) for a given curve, the change of the chemical departure value almost does not affect the reduction factor until it increases at a value; (II) for huge gaps $\log_{10}(gap/T)\geq 1$, as soon as $\delta\mu > gap$, and until the reduction factor becomes close to 1, the curves are almost parallel. This means that for small chemical departure, the reduction factor can be written as the function of the variable $gap$ only, while for huge gaps it can be written as $R\sim\mathfrak{f}(gap)\mathfrak{g}(gap/\delta\mu)$. These are similar to the net reaction rates presented by Villain and Haensel \cite{vh05}.

\begin{figure}[t]
\setlength{\unitlength}{1mm}
\leavevmode
\hskip  0mm
\includegraphics[width=0.5\textwidth]{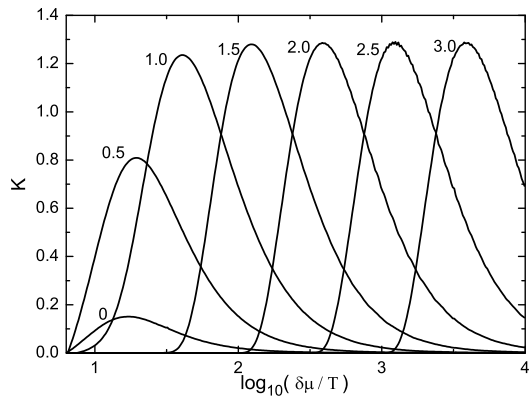}
\caption
{Dependence of $K$ ($K=\partial R/\partial(\log_{10}(\delta\mu/T))$) on the chemical imbalance for various gap's amplitude. The numbers next to the curves give value of $\log_{10}(gap/T)$. Direct reaction with isotropic superfluidity of protons or neutrons. }
\label{2fig}
\end{figure}

Additionally, the behavior of the superfluid suppression on the
neutrino emissivity here is different from that proposed by
Reisenegger \cite{rei97}. He suggested that for $\delta\mu <
\Delta_p + \Delta_n$ the beta reactions were completely suppressed
($R=0$), while for $\delta\mu > \Delta_p + \Delta_n$ the effect of
superfluidity could be neglected ($R=1$). However, as can be seen
from this figure, the reduction factor $R$ doesn't equal to $0$ or
$1$. In other words, we find that in a proper range of the value
$\delta\mu/T$, the neutrino emissivity will be suppressed by the
nucleon superfluidity, but it is not completely suppressed. If
taking the onset of rising R as the threshold point, we find that in
the region of $\log_{10}(gap/T) \leq 1.0$,
$\log_{10}(\delta\mu_{th}/T) = constant$ (approximately 0.78 in Fig.
\ref{2fig}), while for $\log_{10}(gap/T) > 1.0$,
$\delta\mu_{th}\approx gap$. Here $\mu_{th}$ is the value of $\mu$
at the threshold point. Eventually, when and only when the value of
$gap/T$ approaches infinity, $gap/T \rightarrow \infty$, our results
are fully same with that in Reisenegger \cite{rei97}. These are
easily illustrated by plotting $K$ as function of $\delta\mu/T$ in
Fig. \ref{2fig}, where $K$ is the partial derivative of $R$ with
respect to $\log_{10}(\delta\mu/T)$. It is clearly that the curves
in this figure can confirm our above discussions.

The above behavior of superfluid suppression on the neutrino
emissivity can also be seen from all the other comparable figures in
the following.

\begin{figure}[t]
\setlength{\unitlength}{1mm}
\leavevmode
\hskip  0mm
\includegraphics[width=0.5\textwidth]{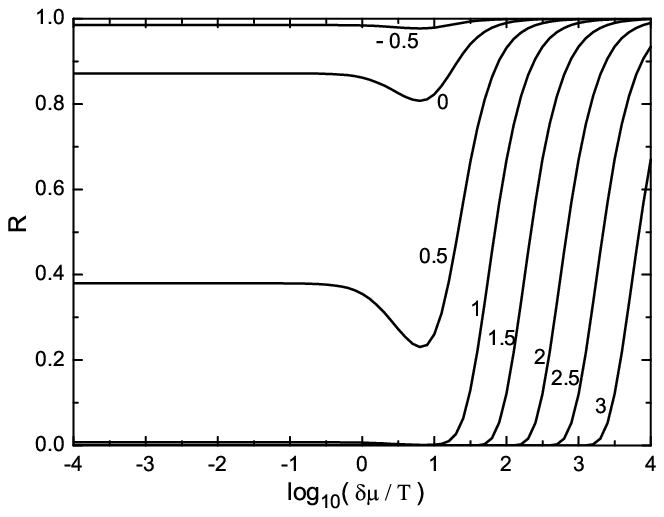}
\caption
{Same as in Fig. (\ref{1fig}) but with anisotropic superfluidity type B of the neutrons. }
\label{3fig}
\end{figure}

\begin{figure}[t]
\setlength{\unitlength}{1mm}
\leavevmode
\hskip  0mm
\includegraphics[width=0.5\textwidth]{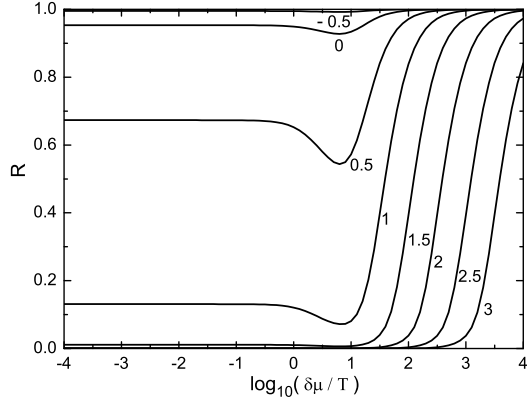}
\caption
{Same as in Fig. (\ref{1fig}) but with anisotropic superfluidity type C of the neutrons. }
\label{4fig}
\end{figure}

Figs. \ref{3fig} and \ref{4fig} show the dependence of the reduction factor of Durca reaction on dimensionless chemical departure and dimensionless gap amplitude with anisotropic superfluidity of B and C, respectively. Compared with Fig. \ref{1fig}, we can see that the reduction factors can either increase (type C) or decrease (type B) in respect to that of type A, which means that the maximal value of the gap on the Fermi surface is important since the value for case C is larger than for case A, while for case B is smaller.
%
%%%%%%%%%%%%%%%%%%%%%%%%%%% FIGURE 1  %%%%%%%%%%%%%%%%%%%%%%%%%%%%%
%%%%%%%%%%%%%%%%%%%%%%%%%%%%%%%%%%%%%%%%%%%%%%%%%%%%%%%%%%%%%%%%
%
%
%%%%%%%%%%%%%%%%%%%%%%%%%%% FIGURE 2  %%%%%%%%%%%%%%%%%%%%%%%%%%%%%
%%%%%%%%%%%%%%%%%%%%%%%%%%%%%%%%%%%%%%%%%%%%%%%%%%%%%%%%%%%%%%%%
%
%
%%%%%%%%%%%%%%%%%%%%%%%%%%% FIGURE 3  %%%%%%%%%%%%%%%%%%%%%%%%%%%%%
%%%%%%%%%%%%%%%%%%%%%%%%%%%%%%%%%%%%%%%%%%%%%%%%%%%%%%%%%%%%%%%%
%
%
%%%%%%%%%%%%%%%%%%%%%%%%%%% FIGURE 4  %%%%%%%%%%%%%%%%%%%%%%%%%%%%%
%%%%%%%%%%%%%%%%%%%%%%%%%%%%%%%%%%%%%%%%%%%%%%%%%%%%%%%%%%%%%%%%
%

%%%%%%%%%%%%%%%%%%%%%%%%%%%%%%%%%%%%%%%%%%%%%%%%%%%%%%%%%%%%%%%%%%%%%%%%%%%%%%%%
\section{Modified Urca processes}
%%%%%%%%%%%%%%%%%%%%%%%%%%%%%%%%%%%%%%%%%%%%%%%%%%%%%%%%%%%%%%%%%%%%%%%%%%%%%%%%
As mentioned in the previous section, the direct Urca process in $npe$ matter is allowed at densities several times the nuclear density. Then in smaller densities the modified Urca process becomes the main neutrino reaction. Reaction (\ref{eq:Murca}) differs from reaction (\ref{eq:Durca}) by the presence of an additional nucleon to ensure momentum conservation. The modified Urca process will be labeled by upperscripts (MN), where $N=n$ indicates the neutron branch of reaction (\ref{eq:Murca}) and $N=p$ indicates the proton branch. Similar to above section, we give the expressions of the neutrino emissivities for the direct and inverse Murca processes out of equilibrium in superfluid nuclear matter,
\begin{eqnarray}
&& Q_n^{(MN)}(\xi,v_j)={4\pi \over {2(2 \pi)^{14}}} T^8\left[ \prod_{j=1}^5 \int d\Omega_j\right]\delta(\mathbf{P}_f-\mathbf{P}_i)|M_{fi}|^{2} \prod_{j=1}^5 p_{F_j} m_j^\ast\int_0^\infty \dd x_\nu \; x_\nu^3
\nonumber \\
&&    \int_{-\infty}^{+\infty} \dd x_1\dd x_2 \dd x_3 \dd x_4\dd x_5\; f(z_1)f(z_2) \;(1-f(z_3))
     \;(1-f(z_4))(1-f(x_5))\delta(z_3+z_4+x_5-z_1-z_6+x_\nu + \xi),
\label{eq:QMNn}
\end{eqnarray}
\begin{eqnarray}
&& Q_p^{(MN)}(\xi,v_j)={4\pi \over {2(2 \pi)^{14}}} T^8\left[ \prod_{j=1}^5 \int d\Omega_j\right]\delta(\mathbf{P}_f-\mathbf{P}_i)|M_{fi}|^{2} \prod_{j=1}^5 p_{F_j} m_j^\ast
\int_0^\infty \dd x_\nu \; x_\nu^3
\nonumber \\
&&    \int_{-\infty}^{+\infty} \dd x_1\dd x_2 \dd x_3 \dd x_4\dd x_5\;(1-f(z_1))(1-f(z_2)) f(z_3) \;f(z_4)f(x_5)
\delta(z_3+z_4+x_5-z_1-z_2-x_\nu - \xi).
\label{eq:QMNp}
\end{eqnarray}
where $j=1,2,3,4$ correspond to $n,N_i,p,N_f$ ($N_i$ and $N_f$ refer to the initial and finial spectator nucleons), while $j=5$ corresponds to the electron. The remaining parameters are the same as in Durca case. The total neutrino emissivity of one branch of Murca processes is $Q^{(MN)} =Q_n^{(MN)} +Q_p^{(MN)}$.

If none of the nucleons is superfluid, we have no necessary to make distinction between the two branches in the phase-space integral and $z_j = x_j$ respectively. Under the condition of chemical equilibrium, $\xi=0$, the neutrino emissivity of direct and inverse Murca processes without nucleon superfluidity are the same. The total neutrino emissivity of one branch of Murca processes have been calculated by \cite{fm79},
\begin{eqnarray}
&&    Q_{0}^{(MN)}  =  {1 \over {(2 \pi)^{14}}} T^8 A_{MN}I_{0}^{(MN)} |M_{fi}|^{2} \prod_{j=1}^5 p_{F_j} m_j^\ast
\label{eq:Decompmur} \\
&&   A_{MN} = 4\pi \left[ \prod_{j=1}^5 \int d\Omega_j\right]\delta(\mathbf{P}_f-\mathbf{P}_i),
\label{eq:Amur} \\
&&   I_{0}^{(MN)} = \int_0^\infty \dd x_\nu \; x_\nu^3
       \left[ \prod_{j=1}^5 \int_{-\infty}^{+\infty}
       \dd x_j \; f_j \right]
       \delta \left( \sum_{j=1}^5 x_j-x_\nu \right),
\label{eq:IMN}
\end{eqnarray}
Here $f_j=(1+e^{x_j})^{-1}$ is the Fermi-Dirac distribution function. The quantity $|M_{fi}|^2$ appear to be constant and can be taken out the integral. The detailed calculations refer to \cite{ykgh01}. we have
\begin{equation}
     A_{Mn} =\frac{2\pi (4\pi)^4}{\pFn^3}, \quad A_{Mp} =\frac{2(2\pi)^5}{\pFn \pFp^3 \pFe}(\pFe + 3 \pFp -\pFn), \,
     \quad
     I_{0}^{(MN)}=\frac{11513\pi^8}{60480}.
\end{equation}

\subsection{Non-equilibrium without superfluidity}
The neutrino emissivity of non-equilibrium modified Urca process without nucleon superfluidity has been studied by some authors (\cite{h92,rei95}). When the matter is out of beta equilibrium, $\delta\mu \neq 0$, it opens additional volume in the phase-space which results in an increase of the neutrino emissivity. It can be written as:
\begin{equation}
Q_{MN}(\xi)={1 \over {2(2 \pi)^{14}}} T^8 A_{MN}I_{MN}(\xi) |M_{fi}|^{2} \prod_{j=1}^5 p_{F_j} m_j^\ast,
\label{eq:QMNNOSF}
\end{equation}
with
\begin{eqnarray}
&& I_{MN}(\xi)=\int_0^\infty dx_\nu x_\nu^3 \left[J(x_\nu - \xi) + J(x_\nu + \xi)\right],
\label{eq:IMNNOSF}\\
&& J(x_\nu - \xi)= \left[ \prod_{j=1}^5 \int_{-\infty}^{+\infty}
       \dd x_j \; f_j \right]\delta(\sum_{j=1}^5 x_j - x_\nu + \xi)
        =\frac{\pi^2 +(x_\nu - \xi)^2}{2(e^{x_\nu - \xi} +1)},
\label{eq:Jxminus}\\
&& J(x_\nu + \xi)= \left[ \prod_{j=1}^3 \int_{-\infty}^{+\infty}
       \dd x_j \; f_j \right]\delta(\sum_{j=1}^5 x_j - x_\nu -\xi)
        =\frac{\pi^2 +(x_\nu + \xi)^2}{2(e^{x_\nu + \xi} +1)}.
\label{eq:Jxplus}
\end{eqnarray}
Then the effect of the equilibrium departure on the neutrino emissivity of Murca process can be presented as:
\begin{eqnarray}
&& \frac{Q_{MN}(\xi)}{Q_{0}^{(MN)}}=\frac{I_{MN}(\xi)}{2I_{0}^{(MN)}}.
\end{eqnarray}

According to \cite{rei95},
\begin{eqnarray}
&& Q_{MN}(\xi)=Q_{0}^{(MN)}F_M(\xi) \\
&& F_M(\xi) =1+\frac{22020\xi^2}{11513\pi^2}+\frac{5670\xi^4}{11513\pi^4}+\frac{420\xi^6}{11513\pi^6}+\frac{9\xi^8}{11513\pi^8}.
\end{eqnarray}
Thus
\begin{equation}
I_{MN}(\xi)=\frac{11513\pi^8}{60480}\left (1+\frac{22020\xi^2}{11513\pi^2}+\frac{5670\xi^4}{11513\pi^4}+\frac{420\xi^6}{11513\pi^6}+\frac{9\xi^8}{11513\pi^8} \right)
\end{equation}
%
%%%%%%%%%%%%%%%%%%%
\subsection{Non-equilibrium with superfluidity}
%%%%%%%%%%%%%%%%%%%
The analysis of the modified Urca process is quit similar with the direct Urca process. The appearance of nucleon superfluidity reduces the available phase-space which suppresses the neutrino emissivity, $z_j\neq x_j$. We label $R^i_{MN}(\xi,v_j)$ as the superfluid reduction factor. We can write
\begin{equation}
Q_{MN}(\xi,v_j)= Q_{MN}(\xi) R^i_{MN}(\xi,v_j).
\end{equation}
Notice that for $T > T_c$ (the critical temperature of superfluid nucleon) $R^i_{MN}(\xi,v_j) = R^i_{MN}(\xi,0) = 1$, and for $T < T_c$ $R^i_{MN}(\xi,v_j) < 1$. The total neutrino emissivity $Q^{(MN)}(\xi,v_j) =Q_n^{(MN)}(\xi,v_j) +Q_p^{(MN)}(\xi,v_j)$ is in the following form:
\begin{equation}
Q_{MN}(\xi,v_j)= {1 \over {2(2 \pi)^{14}}} T^8 A_{MN}I_{MN}(\xi,v_j) |M_{fi}|^{2} \prod_{j=1}^5 p_{F_j} m_j^\ast,
\end{equation}
where $A_{MN}$ is the same with Eq. (\ref{eq:Amur}) and
\begin{equation}
 I_{MN}(\xi,v_j)=\frac{4\pi}{A_{MN}} \left[ \prod_{j=1}^5 \int d\Omega_j\right]\delta(\mathbf{P}_f-\mathbf{P}_i)\int_0^\infty \dd x_\nu \; x_\nu^3
     \left[ \prod_{j=1}^5 \int_{-\infty}^{+\infty}\dd x_j \; f_j \right] \,
        [\delta( \sum_{j=1}^5 z_j-x_\nu+\xi )+ \delta( \sum_{j=1}^5 z_j-x_\nu-\xi )].
\label{eq:IMNSF}
\end{equation}
Thus, the reduction factor of nucleon superfluidity for non-equilibrium Murca process can be presented by
\begin{equation}
R^i_{MN}(\xi, v_j)=\frac{I_{MN}(\xi,v_j)}{I_{MN}(\xi)},
\end{equation}

In the following, we numerically analyze some special cases that may occur in dense matter and give the diagrams to  illustrate the relationship of superfluid reduction factor with the equilibrium departure and the superfluid gap amplitude.

Case 1. the neutron branch with isotropic superfluidity of protons

Since the singlet-state pairing proton gap is isotropic, the angular and energy integrations in Eq. (\ref{eq:IMNSF}) are separated. The angular integrals are the same as in the non-superfluid case, and we get
\begin{equation}
R^{pA}_{Mn}(\xi,v_p)= \frac{I_{Mn}(\xi,v_p)}{I_{Mn}(\xi)}
                 = \frac{1}{I_{Mn}(\xi)}\int_{-\infty}^{+\infty} dx_p f(z_p)(H_1(z_p + \xi)+ H_1(z_p - \xi)),
\label{eq:RpAMn}
\end{equation}
with
\begin{equation}
H_1(x)= \frac{1}{6} \int_0^\infty ds s^3 \frac{s-x}{exp(s-x)-1} [(s-x)^2 +4\pi^2].
\end{equation}
This result is also valid for the case when the neutron superfluidity is of singlet type in the proton branch of Murca processes.

Case 2. the proton branch with any superfluidity of neutrons

Let the protons be normal, but the neutrons be superfluid due to $nn$ Cooper pairing. The integration over solid angles $d\Omega_j$ of all particles but the neutron in Eq. (\ref{eq:IMNSF}) is the same as for non-superfluid case. We get
\begin{equation}
R^{nX}_{Mp}(\xi,v_n)= \frac{I_{Mp}(\xi,v_n)}{I_{Mp}(\xi)}
                 = \frac{1}{I_{Mp}(\xi)}\int_0^1 d\varsigma \int_{-\infty}^{+\infty} dx_n f(z_n)(H_1(z_n + \xi)+ H_1(z_n - \xi)),
\end{equation}
where $\varsigma =\cos\vartheta$, $\vartheta$ is the angle between the quantization axis and the momentum of the superfluid neutron and $H_1(x)$ is the same as in Eq. (\ref{eq:RpAMn}).

Case 3. the neutron branch with isotropic superfluidity of neutrons

Now consider the neutron branch with isotropic superfluidity of neutrons. Let $j=1, 2, 3$ in Eq. (\ref{eq:IMNSF}) refer to neutron. Integrating over $x_4$ and $x_5$, we get
\begin{equation}
R^{nA}_{Mn}(\xi,v_n)= \frac{I_{Mn}(\xi,v_n)}{I_{Mn}(\xi)}
                 = \frac{1}{I_{Mn}(\xi)}\prod_{j=1}^3\int_{-\infty}^{+\infty} dx_j f(z_j)(H_2(z_1+z_2+z_3 + \xi)+ H_2(z_1+z_2+z_3 - \xi)),
\end{equation}
\begin{equation}
H_2(x)= \int_0^\infty ds s^3 \frac{s-x}{exp(s-x)-1}.
\end{equation}

\begin{figure}[t]
\setlength{\unitlength}{1mm}
\leavevmode
\hskip  0mm
\includegraphics[width=0.5\textwidth]{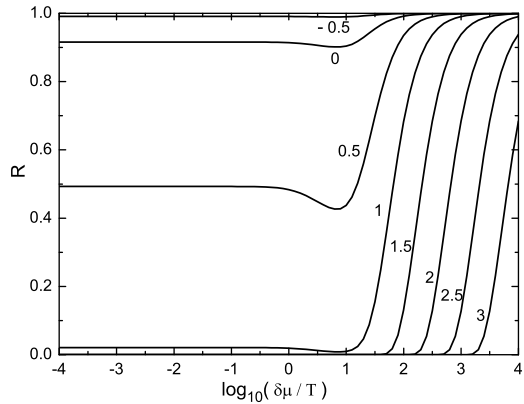}
\caption
{Dependence of the reduction factor on the chemical imbalance for various gap amplitude. The numbers next to the curves give value of $\log_{10}(gap/T)$. Modified neutron branch with isotropic superfluidity of protons. }
\label{5fig}
\end{figure}

\begin{figure}[t]
\setlength{\unitlength}{1mm}
\leavevmode
\hskip  0mm
\includegraphics[width=0.5\textwidth]{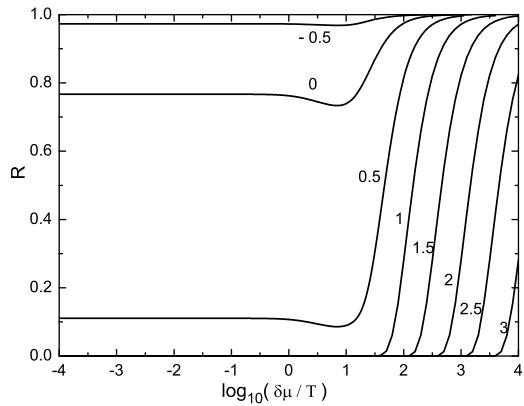}
\caption
{Same as in Fig. (\ref{5fig}) but for Modified neutron branch with isotropic superfluidity of neutrons. }
\label{6fig}
\end{figure}

Employing the numerical calculations similar with that done in \cite{vh05}, we give the results for Case 1 and Case 3 in Figs. \ref{5fig} and \ref{6fig} respectively. We need not give the similar graph for Case 2. The result of Case 2 with isotropic superfluidity of neutrons are same with Case 1. Comparing Figs. \ref{5fig} and \ref{6fig}, we can see that the impact with superfluid spectator nucleons is much larger than that with superfluid non-spectator nucleons. Based on the similar reasons with that presented in \cite{vh05}, the cases with triplet-state nucleon superfluidity have been neglected.

%%%%%%%%%%%%%%%%%%%%%%%%%%% FIGURE 5  %%%%%%%%%%%%%%%%%%%%%%%%%%%%%
%%%%%%%%%%%%%%%%%%%%%%%%%%%%%%%%%%%%%%%%%%%%%%%%%%%%%%%%%%%%%%%%
%
%%%%%%%%%%%%%%%%%%%%%%%%%%% FIGURE 6  %%%%%%%%%%%%%%%%%%%%%%%%%%%%%
%%%%%%%%%%%%%%%%%%%%%%%%%%%%%%%%%%%%%%%%%%%%%%%%%%%%%%%%%%%%%%%%
%
%%%%%%%%%%%%%%%%%%%%%%%%%%%%%%%%%%%%%%%%%%%%%%%%%%%%%%%%%%%%%%%%%%%%%%%%%%%%%%%%
\section{Discussion and Conclusion}
\label{5}
%%%%%%%%%%%%%%%%%%%%%%%%%%%%%%%%%%%%%%%%%%%%%%%%%%%%%%%%%%%%%%%%%%%%%%%%%%%%%%%%
We apply the methods of Villain and Haensel in paper \cite{vh05} to calculate the effects of nucleon superfluidities on the neutrino emissivities of direct and modified Urca processes out of equilibrium, while they calculated the corresponding net reaction rates. Although the numerical methods are similar, there are two differences in calculating neutrino emissivities and net reaction rates which can be seen easily by camparing Eq. (\ref{eq:IMNSF}) and Eq. (25) of paper \cite{vh05}, for example. Firstly, the integral $\int_0^\infty \dd x_\nu \; x_\nu^3$ takes the place of $\int_0^\infty \dd x_\nu \; x_\nu^2$. Secondly, the two delta functions add together in Eq. (\ref{eq:IMNSF}) while subtract each other in Eq. (25) of previous paper. So the calculations of the neutrino emissivities are necessary. We here give the graphical overview of those results for
\begin{description}
\item[-] Durca reaction with superfluidity of each type;
\item[-] Murca reaction with isotropic superfluidity of the
non-spectator nucleon (e.g. proton for the neutron branch);
\item[-] Murca reaction with isotropic superfluidity of the spectator
nucleon (e.g. neutron for the neutron branch).
\end{description}

 As shown in calculations, the appearance of nucleon superfluid reduces the neutrino emissivity by a factor that depends on the value of $\Delta/T$ and of $\delta\mu/T$, the types of the beta process (direct or modified Urca) and of superfluidity. When the quantity $\delta\mu/T$ is relatively small, the superfluid reduction factors don't change with the value of $\delta\mu/T$ and become the function of only one variable $\Delta/T$. It should be similar with that under the condition of beta equilibrium. However, the violation of superfluid suppression on neutrino emissivity occurs when the chemical departure reaches a threshold value. The threshold tends to be the result presented by Reisenegger (\cite{rei97}) only when the energy gap of superfluidity is large enough.

 It's well-known that neutrino emissivity plays a key role on the cooling of neutron stars (\cite{pa92,yp04}). we recognize that although the superfluid gap suppresses dramatically the neutrino emissivity in dense matter, it is possible that the chemical imbalance of beta reaction large enough will break the suppression. Meanwhile, the imbalance contributes chemical heat to neutron star core. Along with the results of reaction rates calculated by Villain and Haensel (\cite{vh05}), our results are required for numerical simulation of neutron star cooling. We have the primary investigations on this problem. It will be reported in our another paper.

%%%%%%%%%%%%%%%%%%%%%%%%%%%%%%%%%%%%%%%%%%%%%%%%%%%%%%%%%%
\section*{Acknowledgments}
%%%%%%%%%%%%%%%%%%%%%%%%%%%%%%%%%%%%%%%%%%%%%%%%%%%%%%%%%%
The authors are grateful to
K.S. Cheng for discussions,
to L. Villain for sending us their subroutines which calculate reaction rates.
This research was supported by NFSC under Grants No 10773004.

%%%%%%%%%%%%%%%%%%%%%%%%%%%%%%%%%%%%%%%%%%%%%%%%%%%%%%%%%%


\begin{thebibliography}{999}

\bibitem{lpph91}
J. M. Lattimer, C. J. Pethick, M. Prakash and P. Haensel,
Phys. Rev. Lett. \textbf{66}, 2701 (1991).
\bibitem{h92}
P. Haensel,
Astron. Astrophys. \textbf{262}, 131 (1992).
\bibitem{rei95}
A. Reisenegger,
Astrophys. J. \textbf{442}, 749 (1995).
\bibitem{rei97}
A. Reisenegger,
Astrophys. J. \textbf{485}, 313 (1997).
\bibitem{fr05}
R. Fern\'andez, and A. Reisenegger,
Astrophys. J. \textbf{625}, 291 (2005).
\bibitem{jrf06}
P. Jofr\'e, A. Reisenegger, and R. Fern\'andez,
Phys. Rev. Lett. \textbf{97}, 131102 (2006).
\bibitem{us96}
V. A. Urpin, and D. A. Shalybkov,
Mon. Not. Astron. Soc. \textbf{281}, 145 (1996).
\bibitem{rg92}
A. Reisenegger, and P. Goldreich,
Astrophys. J. \textbf{395}, 240 (1992).
\bibitem{ykgh01}
D. G. Yakovlev, A. D. Kaminker, O. Y. Gnedin, and P. Haensel,
Phys. Rep. \textbf{354}, 1 (2001).
\bibitem{vh05}
L. Villain, and P. Haensel,
Astron. Astrophys. \textbf{444}, 539 (2005).
\bibitem{ls01}
U. Lombardo and H.-J. Schulze,
in {\it Physics of Neutron Star Interiors},
Springer Lecture Notes in Physics
(Springer, Berlin),
Eds. D. Blaschke, N. K. Glendenning, and A. Sedrakian,
v.~\textbf{578}, p.~30 (2001)
\bibitem{ly94}
K. P. Levenfish, and D. G. Yakovlev,
Astron.\ Lett. {\bf 20} 43 (1994)

\bibitem{lp80}
E. M. Lifshitz and L. P. Pataevskii,
in {\it Statistical Physics}, Part 2,
(Pergamon Press, Oxford, 1980).
\bibitem{tamagaki70}
   Tamagaki~R,
%   Superfluid state in neutron star matter. I. -- Generalized Bogoliubov
%   transformation and existence of $^3P_2$ gap at high density.
   {\it Prog.\ Theor.\ Phys.\ } {\bf 44} 905 (1970)

\bibitem{yls99}
D. G. Yakovlev, K. P. Levenfish, and Yu. A. Shibanov,
Physics-Uspekhi {\bf 42}, 737 (1999).
\bibitem{st83}
S.L. Shapiro , and S.A. Teukolsky, 1983,
in {\it Black Holes, White Dwarfs and Neutron Stars},
   (Wiley-Interscience, New-York, 1983)
\bibitem{fm79}
B. L. Friman, and O. V. Maxwell,
Astrophys. J. {\bf 232}, 541 (1979).

\bibitem{yl95}
    D. G. Yakovlev, and K. P. Levenfish,
%  Modified URCA process in neutron star cores.
    Astron.\ Astrophys. {\bf 297} 717 (1995)
\bibitem{pa92}
   D.Page,  J. H.Applegate,
% The cooling of neutron stars by the direct Urca process
   Astrophys.\ J.\ Lett.\  {\bf 394} L17 (1992)
\bibitem{yp04}
D. G. Yakovlev and C. J. Pethick,
Ann. Rev. Astron. Astrophys. {\bf 42}, 169 (2004).




\end{thebibliography}
\end{document}